\documentclass[11pt]{article}
\usepackage{moriond,epsfig}

\def\Journal#1#2#3#4{{#1} {\bf #2}, #3 (#4)}

\def\PRL{\em Phys. Rev. Lett.}
\def\PRB{{\em Phys. Rev.} B}

\def\NAT{{\em Nature}}
\def\pss{{\em phys. stat. sol.} (b)}
\def\epjb{{\em Eur. Phys. J.} B}

\def\be{\begin{equation}}
\def\ee{\end{equation}}
\def\bea{\begin{eqnarray}}
\def\eea{\end{eqnarray}}

\begin{document}
\title{DYNAMICS OF COULOMB--CORRELATED ELECTRON--HOLE PAIRS IN 
DISORDERED SEMICONDUCTOR NANOWIRES}

\author{\underline{I. VARGA}$^{1,2}$, C. SCHLICHENMAIER$^1$, 
T. MEIER$^1$, K. MASCHKE$^3$,\\ P. THOMAS$^1$ and 
S.W. KOCH$^1$}

\address{$^1$Fachbereich Physik und Wissenschaftliches Zentrum 
f\"ur Materialwissenschaften, Philipps Universit\"at,
Marburg, D-35032 Marburg, Germany}
\address{$^2$Elm\'eleti Fizika Tansz\'ek, Fizikai Int\'ezet,
Budapesti M\H uszaki \'es Gazdas\'agtudom\'anyi Egyetem, 
H-1521 Budapest, Hungary}
\address{$^3$Institut de Physique Appliqu\'ee,
\'Ecole Polytechnique F\'ed\'erale de Lausanne, CH-1015 Lausanne,
Switzerland}

\maketitle\abstracts{The dynamics of optically generated electron-hole 
pairs is investigated in a disordered semiconductor nanowire. The 
particle pairs are generated by short laser pulses and their dynamics 
is followed using the Heisenberg equation of motion. Is is shown that 
Coulomb--correlation acts against localization in the case of the 
two--interacting particles (TIP) problem. Furthermore, currents are 
generated using a coherent combination of full-gap and half-gap pulses. 
The subsequent application of a full-gap pulse after time $\tau$ produces 
an intraband echo phenomenon $2\tau$ time later. The echo current is shown to 
depend on the mass ratio between the electrons and the holes}

\section{Introduction}

Optically generated carriers in a semiconductor environment produce an
exciting scenario to study fast quantum coherent phenomena. The presence of
disorder and the interaction between the carriers as well as with external
fields produces a number of interesting phenomena~\cite{rev}. The 
prospectives added by the recent fabrication of rings on a nanoscopic range
\cite{nano} may add a further boost to the theoretical investigation of 
low--dimensional small semiconductor systems.

In this contribution we present two examples where the time evolution of the
electron--hole pairs provides interesting insight into the combined effect
of disorder and interaction. In the first one the spreading of initially
localized electron--hole wave packets are shown to expand to considerably
larger extent as compared to the size of single particle localization. 
In the second example we show how the phase coherent dynamics results in a 
new type of echo phenomenon discussed only recently.

\section{The model and the equation of motion}

The Hamiltonian of the two--band semiconductor subject to the external
laser field represented by an electric field $E(t)$ reads
\be
H=\sum_{\lambda=e,h}H_{\lambda}+H_I+H_C, 
\label{tot}
\ee
where the two bands, the conduction band with electrons, $H_e$, and the
valence band with holes, $H_h$, are both written in the 
tight--binding approximation
\be
H_{\lambda} = \sum_{i}\varepsilon_i^{\lambda}\sigma_{ii}^{\lambda\lambda}
-J^{\lambda}\sum_i(\sigma_{i,i+1}^{\lambda\lambda} +
                   \sigma_{i,i-1}^{\lambda\lambda}).
\label{tb}
\ee
The interaction with the laser is kept within the dipole approximation 
\be
H_I = -E(t)d, \qquad\mbox{with}\qquad
d=-e\sum_{\lambda \lambda' i}(R_i\delta_{\lambda\lambda'}+
r_{\lambda \lambda'}) \sigma_{ii}^{\lambda \lambda'}
\label{dip}
\ee
and the Coulomb interaction within the monopole--monopole approximation
\be
H_C = \frac{1}{2}\sum_{ij}(\sigma_{ii}^{ee}-\sigma_{ii}^{hh})V_{ij}
(\sigma_{jj}^{ee}-\sigma_{jj}^{hh}).
\label{eq:c}
\ee
In the above equations we used 
$\sigma_{ij}^{\lambda\lambda'}={a_{\lambda i}^{\dag}}a_{\lambda' j}$, if
$a_{\lambda i}^{\dag}$ ($a_{\lambda i}$) creates (annihilates) a particle
in the band $\lambda$ ($e$ or $h$) at site $i$. Hence in Eq.~(\ref{eq:c}) 
the electron densities are described using 
$n_{ij}^e=\langle\sigma_{ij}^{ee}\rangle$ and the hole densities using 
$n_{ij}^h=\delta_{ji}-\langle\sigma_{ji}^{hh}\rangle$. The interband 
coherences (the pair amplitudes) are 
$p_{ij}=\langle\sigma_{ij}^{eh}\rangle$. The interaction between the 
carriers is assumed to have the regularized form of 
$V_{ij}=V_0\,a/(|i-j|a+a_0)$, where $a$ is the lattice constant, $a_0$ and 
$V_0$ characterize the strength of the interaction ensuring a finite value 
for the excitonic binding energy. For both of our studies they were fixed in
order to obtain an exciton binding energy typical for quasi--onedimensional 
GaAs quantum wires.

In our study disorder entered by choosing the on--site energies 
$\varepsilon_i^{\lambda}$ from a flat distribution of width $W$. These
random values at each site for different bands may be anti-correlated
if the energy separations of the isolated sites are constant and correlated
if the averages of the site energies coincide.

The time evolution of the various components of 
$\langle\sigma_{ij}^{\lambda\lambda'}\rangle$ 
are obtained via the solution of the Heisenberg equations of motion 
(using $\hbar=1$)
\be
i\partial_t\sigma_{ij}^{\lambda \lambda'}=
\left [\sigma_{ij}^{\lambda \lambda'},H\right ]
\ee
In our first example we will simplify the equation assuming low excitation
intensity. Thus linear response in the external laser field and henceforth 
the Hartree--Fock approximation for the particle--particle interaction are
valid. This results for the interband coherences in
\be
i\partial_t p_{ij}= \left(\epsilon_j^e+\epsilon_i^h-V_{ij}\right )-
J^h\left(p_{i,j-1}+p_{i,j+1}\right)+J^e\left(p_{i-1,j}+p_{i+1,j}\right)-
\mu_jE(t)\delta_{ij}
\label{eq:pij}
\ee
where $\mu_j$ is the polarizability at site $j$. The intraband coherences,
in this case, are calculated using the sum rules, 
$n^e_{ij}=\sum_k p_{kj}p_{ki}^*$ and $n^h_{ij}=\sum_k p_{jk}p_{ik}^*$ valid
in this low excitation limit.

The second problem requires the external field to be handled 
nonperturbatively. Therefore as a first step we have omitted the 
particle--particle interaction or only kept it at the level similar to
that presented in Eq.~(\ref{eq:pij}). In this case the term containing
the external field would be 
$eE(t)\left [(R_i-R_j)p_{ij}+r_{eh}(n^e_{ij}+n^h_{ji}-\delta_{ij})\right ]$. 
The intraband amplitudes, $n^e_{ij}$ and $n^h_{ij}$ are obtained in a similar 
fashion and have to be solved simultaneously.

\section{Two interacting particles}

\begin{figure}
\begin{center}
\epsfig{figure=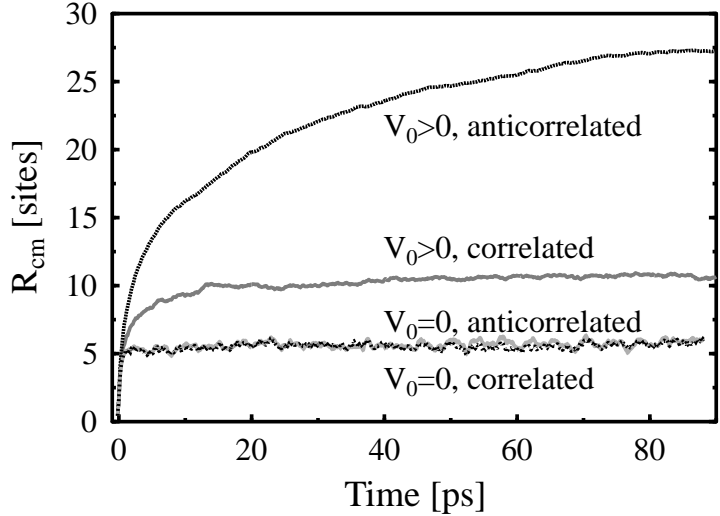,height=2.7in}
\end{center}
\caption{Center of mass width of a two--particle wave packet for
correlated and anti-correlated disorder. $J^e=-J^h=20$meV, $W^e=W^h=4J^e$. 
The curves have been obtains after averaging over 20 samples.
\label{f:tip}}
\end{figure}

The time evolution of an initially localized excitation has been followed
via the integration of the equations of motion (\ref{eq:pij}) for the 
interband coherences, $p_{ij}$. The extension of the wave packet $p_{ij}$ 
was characterized, among others, by the center of mass width, 
$R_{cm}^2=\frac{1}{2}\sum_{ij}(i+j)^2|p_{ij}|^2$.
A typical evolution of $R_{cm}$ is given in Fig.~\ref{f:tip}. Here we used
$N=240$ sites, $J^e=-J^h=20$meV and very strong disorder, $W^h=W^e=4J^e$. 

The case of anti-correlated disorder seems to be much more favorable in order
to produce reduced localization of the wave packets as is shown in 
Fig.~\ref{f:tip}. The numerically obtained time functions may be fitted
with high accuracy to a phenomenological form of 
$R(t)=((Dt)^{-1/2}+\xi_{\infty}^{-1})^{-1}$ allowing a diffusion constant 
$D$ and an extrapolated infinite time localization length $\xi_{\infty}$
to be obtained from the fitting procedure~\cite{murcia}. 

\section{Coherent control and current echo}

In our other example we show how the rephasing process of {\it intra}band
excitations after two subsequent laser pulses with a time delay of $\tau$ 
can be investigated. This is in contrast to the photon echo which is 
similarly seen in the {\it inter}band excitation. The observable showing 
such an echo phenomenon is the total intraband current. 

The echo phenomenon has been first suggested
using a voltage pump~\cite{echo0} for noninteracting particles. In the same
scheme it has been shown that the many particle interaction reduces its
height but does not destroy the current echo~\cite{Sauter}. Recently 
the use of coherent control provides a further possibility to generate such 
an echo phenomenon~\cite{stippler}. 

A current in the system (\ref{tot})--(\ref{eq:c}) may be 
generated using a special excitation in which the initial 
populations are produced via transitions that have different initial 
propagation into different spatial directions, i.e. a spontaneous current 
appears. This scheme is an example of coherent control~\cite{cc}. It is 
realized using a pulse with mean frequency in resonance with transitions 
in the optical continuum and another pulse having only half that frequency,
$E(t)=\mbox{Re}\left\{E_1 e^{i\omega t+\phi_1} + 
E_2 e^{i\omega t/2+\phi_2}\right\}$.
The resulting current can be calculated using the equation of motion,
${\cal J}={\dot d}=i[H,d]$ which results in a sum of two 
terms, an intraband current, $\langle{\cal J}_{intra}\rangle$, and an 
interband current, $\langle{\cal J}_{inter}\rangle$. Here we give the 
former as 
$\langle{\cal J}_{intra}\rangle\sim J^e \sum_i \mbox{Im}(n_{i+1,i}^e)
-J^h\sum_i\mbox{Im}(n_{i+1,i}^h)$.

The renewed excitation of the system with a laser pulse after time $\tau$ 
causes phase conjugation. As it turns out we only need a light field with 
mean frequency $\omega$ for the second excitation. This phase conjugation 
leads to a collective rephasing of the electron and the hole intraband 
coherences, $n_{ij}^{e,h}$, at times~\cite{Christoph}
\be
T_e=\tau+\frac{J^h}{J^e} \tau\qquad\mbox{and}\qquad
T_h=\tau+\frac{J^e}{J^h} \tau.
\ee
This is when the echo occurs in the conduction (valence) band for electrons 
(holes). Hence in the case when the holes are heavier than the electrons,
$J^e>J^h$, the echo signal contains two peaks: one before $t=2\tau$
for the electrons and another one after that for the holes. 

\begin{figure}
\begin{center}
\epsfig{figure=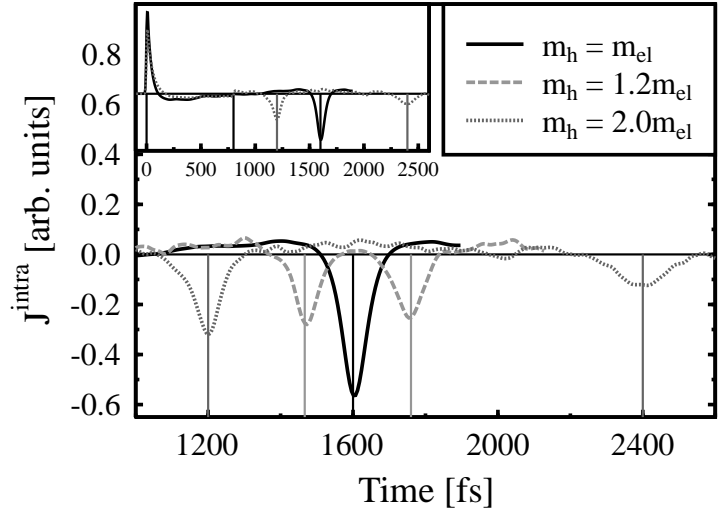,height=2.7in}
\end{center}
\caption{The echo time dependence on the effective mass of the particles.
The inset shows the full time evolution with the first excitation at $t=0$,
the second excitation at $t=800$fs using only a full gap pulse (producing
a negligible response in the intraband current) and the spontaneous echo
at $t=1600$fs. In the main panel the variation of the echo peak
is shown for various electron--hole mass ratios. The curves have been 
obtained as averages over 128 realizations of the disorder.
\label{f:mass}}
\end{figure}

In Fig.~\ref{f:mass} we give an example when 
the second excitation at $t=800$fs has been simplified for a full gap pulse 
instead of the combination present at $t=0$. That is the reason why we 
see only a minor change in the current response of the system at $t=\tau$.
The double echo peak presented in Fig.~\ref{f:mass} is obtained for a system 
consisting of $N=71$ sites, with $m_e=0.28m_0$, $W^{e,h}=2J^{e,h}$ and
correlated disorder.

Note that the present model shows a photon echo even in the absence of 
disorder, while in that case there is no current echo. This emphasizes the
fundamental difference of these two echo phenomena.

Work is in progress to study the influence of the many--particle interaction
on this phenomenon in the coherent--control scenario.

\section*{Acknowledgments}
This work was supported by the Deutsche Forschunggemeinschaft (DFG) through 
the Quantenkoh\"arenz Schwerpunkt, by the Max-Planck Research prize, the 
Leibniz prize and the Hungarian 
Research Fund (OTKA) under T029813, T032116 and T034832.

\end{document}